\documentclass[fp]{jpsj3}
\usepackage{txfonts}
\usepackage{color}
\usepackage{bm}
\title{Deep Learning the Quantum Phase Transitions in Random Electron Systems:
Applications to Three Dimensions}

\author{ Tomi Ohtsuki$^1$\thanks{ohtsuki@sophia.ac.jp} and Tomoki Ohtsuki$^2 $\thanks{ootsuki\_t@msi.co.jp}}
\inst{
$^1$Physics Division, Sophia University, Chiyoda-ku, Tokyo 102-8554, Japan\\
$^2$ NTT DATA Mathematical Systems Inc, Shinjuku-ku, Tokyo 160-0016, Japan} %

\abst{
Three-dimensional random electron systems undergo quantum phase transitions
and show rich phase diagrams.
Examples of the phases are the band gap insulator, Anderson insulator,
strong and weak topological insulators, Weyl semimetal, and diffusive metal.
As in the previous paper on two-dimensional quantum phase transitions
[J. Phys. Soc. Jpn. {\bf 85}, 123706 (2016)],
we use an image recognition algorithm based on a multilayered convolutional neural network
to identify which phase the eigenfunction belongs to.
The Anderson model for localization-delocalization transition,
the Wilson--Dirac model for topological insulators, and the layered Chern insulator model
for Weyl semimetal are studied.
The situation where the standard transfer matrix approach is not applicable is also treated by
this method.
}

\begin{document}
\maketitle

\section{Introduction}
Random three-dimensional (3D) electron systems show rich insulator and metal phases:
band gap insulator, Anderson insulator,\cite{Anderson58}
strong and weak topological insulators,\cite{Hasan10,Qi11}
Weyl semimetal,\cite{Murakami07,Burkov11} and diffusive metal are examples of the phases.
The appearances of these phases\cite{Schnyder08,Kitaev09,Hasan10,Qi11} are related to
the basic symmetry such as time reversal, spin rotation, chiral, and
particle--hole symmetries. \cite{Wigner51,Dyson61,Dyson62,Zirnbauer96,altland97}

The eigenfunction of each phase exhibits specific features.
In the Anderson insulator,  the eigenfunctions are exponentially localized.
Strong topological insulators (STIs) have two-dimensional (2D) Dirac electrons
on all their surfaces, whereas weak topological insulators (WTIs) have
2D Dirac electrons on specific surfaces determined by weak indices.\cite{Fu07}
In the Weyl semimetal (WSM), surface states that form a Fermi arc appear on surfaces
parallel to the direction of split Weyl nodes.\cite{Okugawa14,Chen15,Liu16}

In the presence of randomness, however, owing to the large amplitude fluctuations
of eigenfunctions, determining the phases from eigenfunctions is
difficult, especially when the system is close to phase boundaries.
Note that topological numbers are usually defined in randomness-free systems via the
integration of the Berry curvature of the Bloch function over the Brillouin zone.
The definition is no longer valid once  randomness is introduced, which
breaks the translational invariance. 
Hence, a new definition of topological numbers in random systems is necessary.\cite{Sbierski14a,Katsura16,Katsura16a}

In our previous paper,\cite{Tomoki16}
we regarded the modulus squared (probability density) of an eigenfunction
as an image, and used image recognition\cite{Obuchi14}  based on deep learning\cite{LeCun15,Silver16}
to determine which phase the eigenfunction belongs to.
Changing the stream of random numbers (MT2023) and their seeds,
we have trained the convolutional neural network (CNN) by preparing a large number
of eigenfunctions belonging to each phase of random 2D electron systems,
and have identified the phase of independently obtained eigenfunctions.
The localization--delocalization transition in the presence of strong spin--orbit coupling
(symplectic class),\cite{Asada02} as well as the Chern insulator--Anderson
insulator transition (unitary class),\cite{Dahlhaus11,Liu16} has been verified.
 
In this paper, we further advance this approach to deal with 3D systems.
3D systems show even richer phases than 2D, such as
strong and weak topological insulators and Weyl semimetals.
We will first study the Anderson-type 3D localization--delocalization
transition.  Then, the phase diagrams of topological insulators are derived,
followed by the demonstration of the phase diagram for the layered Chern insulator that exhibits
3D Chern insulators (CIs), WSM, and diffusive metal (DM) phases.
We will demonstrate that the phase diagram can be obtained even in cases where
the transfer matrix approach\cite{RyuNomura:3DTI,Kobayashi13} is not applicable.

\section{Method}
We adopt a four-weight-layer convolutional neural network that outputs real numbers
$\in [0,1]$.
Each number corresponds to the probability with which the input image belongs to a certain phase. 
In the case of the Anderson transition, the output numbers are two, each corresponding to
the probability of a delocalized/localized phase.
In the case of topological insulators, the output numbers are four, corresponding to
an ordinary insulator (OI), WTI, STI, and DM.
The output numbers are three for the layered Chern insulator, corresponding to
CI, WSM, and DM.

We diagonalize 3D random electron systems on a cubic lattice using the
sparse matrix diagonalization algorithm: Intel MKL/FEAST.
We impose a periodic boundary condition in the $y$-direction,
and integrate the modulus squared of the eigenfunction $\psi(x,y,z)$ to
obtain the input 2D image $F(x,z)$,
\begin{equation}
\label{eq:2dTo3d}
F(x,z)=\int \mathrm{d}y\, |\psi(x,y,z)|^2\,.
\end{equation}
The lengths in the $x$- and $z$-directions are 40,
so the input image is 40$\times$40.
The lengths in the $y$-direction are 40 for the Anderson model and layered
 Chern insulator, and 20 for the topological insulator.

When we calculate the probability of the delocalized/localized phase in the case of the 3D Anderson transition,
the average over 5 samples is
 taken so that we can compare the 3D result with the case of the 2D Anderson transition in our previous paper.\cite{Tomoki16}
In the cases of the topological insulator and layered Chern insulator, owing mainly to the limitation of computational power, we
do not take the sample average.

The four-weight-layer CNN used in this study
is a variant of LeNet \cite{lecun1998gradient} included
in Caffe \cite{jia2014caffe} (with the input size changed to $40\times 40$), which utilizes  the 
rectified linear unit (ReLU) as its activation function.
The network weight parameters to be trained are sampled from the Gaussian distribution of
scale determined by the number of input and output dimensions\cite{glorot2010understanding},
except for the first convolution layer connected to the input image:
in the first convolution layer, we have manually chosen the weight initialization scale to be 100.
This choice of weight initialization scale is due to the fact that we deal with eigenfunctions, the values of which are typically much smaller than those of gray-scale images.

We have used the RMSProp solver\cite{tieleman2012lecture} with the parameters in
the Caffe MNIST example (which is contained as \verb+examples/mnist/lenet_solver_rmsprop.prototxt+ in the Caffe source) as the stochastic gradient descent solver.
During the training, we randomly partition the training data set into 90 and 10\%, and use the latter as the validation set.
The solver performs sufficient iterations so that the validation error becomes stationary.
Further details can be found in our previous paper.\cite{Tomoki16}

\section{Applications to 3D Quantum Phase Transitions}
\subsection{Anderson transition}
We first use the 3D Anderson model of localization and discuss the Anderson transition\cite{Anderson58}.
The Hamiltonian is given by
\begin{equation}
\label{eq:su2Hamiltonian}
H=\sum_{\bm{x}}
v_{\bm{x}} c_{\bm{x}}^\dagger c_{\bm{x}}-
\sum_{\langle \bm{x},\bm{x}'\rangle} c_{\bm{x}}^\dagger c_{{\bm{x}}'}\,,
\end{equation}
where $c_{\bm{x}}^\dagger$ ($c_{\bm{x}}$) denotes the creation (annihilation)
operator of an electron at a site $\bm{x}=(x,y,z)$, and $v_{\bm{x}}$
denotes the random potential at $\bm{x}$.
$\langle \cdots \rangle$ indicates the nearest-neighbor hopping.
We assume a box distribution with
each  $v_{\bm{x}}$ uniformly and independently distributed at the interval
$[-W/2,W/2]$.
At energy $E=0$, i.e., at the band center, the wave functions are delocalized
when $W<W_c$ and the system is a diffusive metal.  For $W>W_c$, the wave functions
are localized and the system is an Anderson insulator (AI).
$W_c$ is estimated to be around $16.54$ by the transfer matrix
method combined with the  finite size scaling analyses.\cite{slevin14}
We note here that the multifractal finite size scaling analyses of wave functions
reproduce the critical behavior of the 3D Anderson model obtained by
the transfer matrix method.\cite{alberto10,alberto11}

\begin{figure}[htbp]
  \begin{center}
    \begin{tabular}{c}
    
     \begin{minipage}{0.45\hsize}
  \begin{center}
   \includegraphics[width=0.95\textwidth]{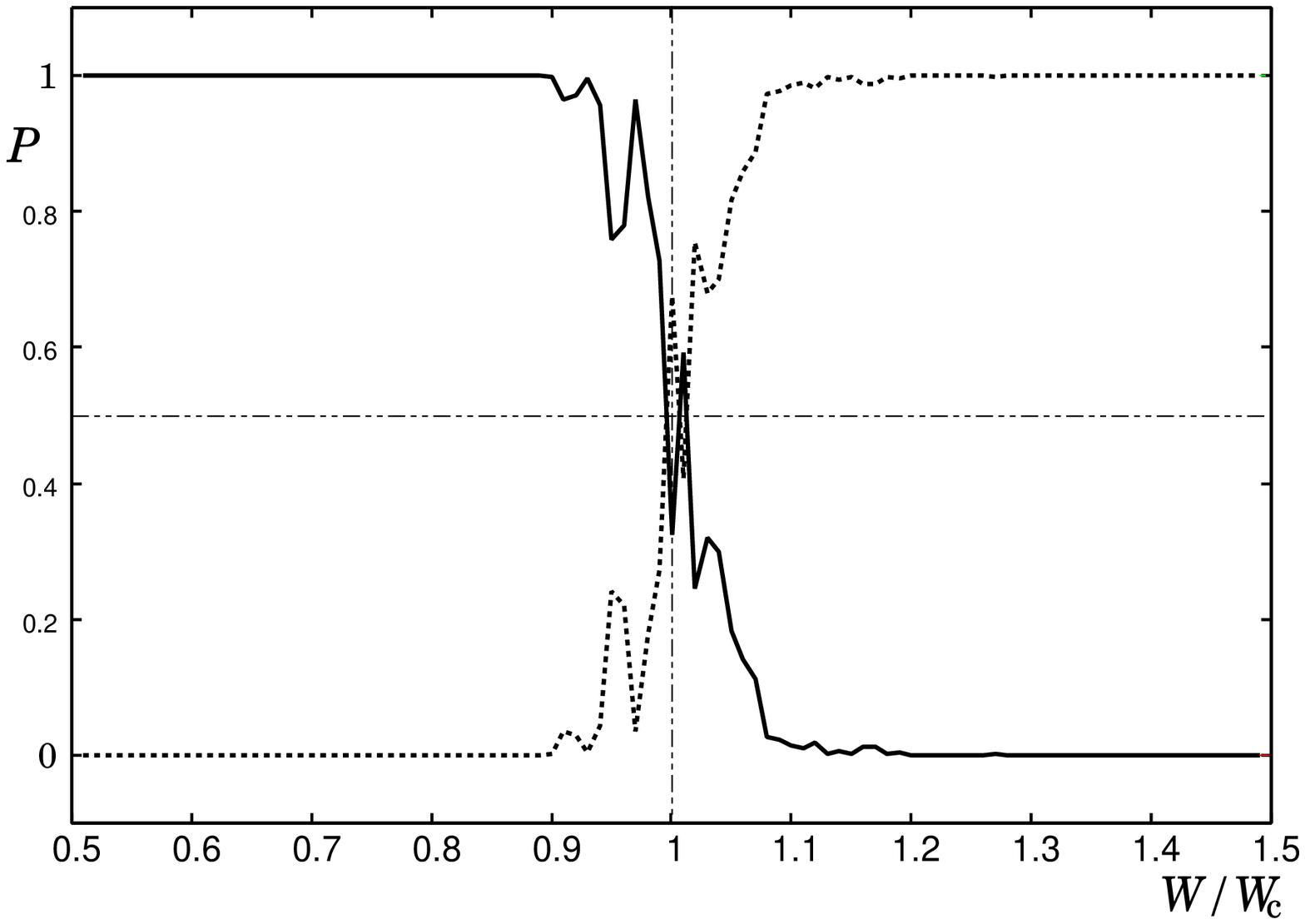}
   \hspace{1.6cm} (a)
     \end{center}
 \end{minipage}
 
 \begin{minipage}{0.45\hsize}
  \begin{center}
   \includegraphics[width=0.95\textwidth]{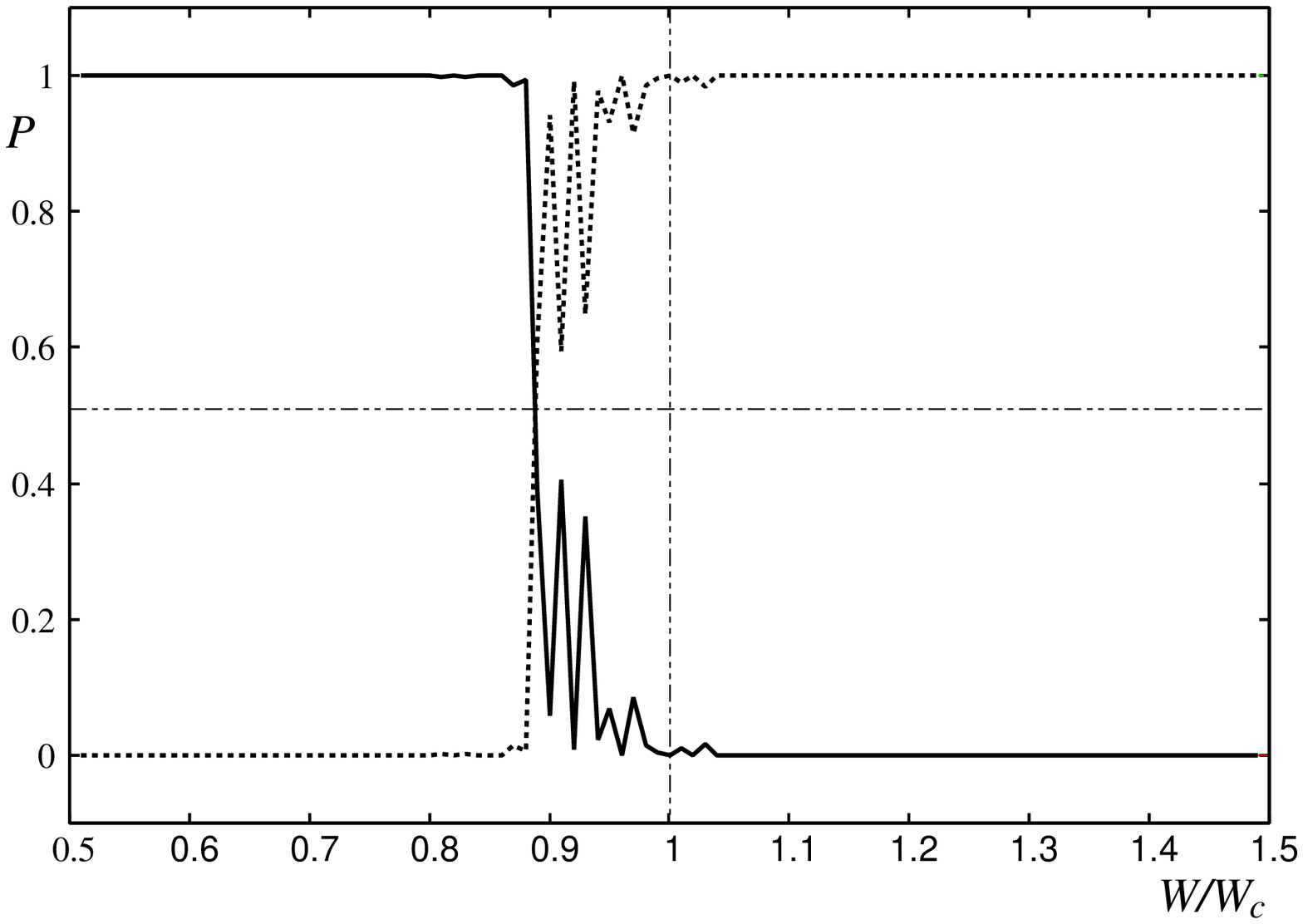}
      \hspace{1.6cm} (b)
     \end{center}
 \end{minipage}
  \end{tabular}
 \caption{Probabilities that the eigenfunction near the band center belongs to the delocalized 
  (solid line) and  localized (dotted line) phases.
 The average over 5 samples is taken.
 Results of learning  (a) the 3D Anderson transition and (b)  2D Anderson transition.
 The dashed dotted lines indicate 50\% probability (horizontal line) and the critical disorder
 (vertical line).}
\label{fig:3DAndersonModel}
\end{center}
\end{figure}

As in the 2D symplectic case\cite{Tomoki16}, we prepared 1000 eigenfunctions
in the diffusive metal regime and 1000 eigenfunctions in the Anderson insulator regime.
We then prepared 100 eigenfunctions with various strengths of
disorder $W$, and let the machine determine whether they are delocalized (DM phase) or localized (AI phase).
Results are shown in Fig.~\ref{fig:3DAndersonModel}(a),
where the probabilities that the eigenfunction belongs to the DM phase $P_\mathrm{DM}$
and to the AI phase $P_\mathrm{AI}=1-P_\mathrm{DM}$ are plotted
as solid and dotted lines, respectively.
A sharp transition from a metal to an insulator is observed near $W_c$.

As a test, the CNN trained for  2D DM--AI transition\cite{Tomoki16} was used to determine the 3D eigenfunctions.
Results are plotted in  Fig.~\ref{fig:3DAndersonModel}(b), where
the transition is qualitatively captured, although the transition is indicated
earlier than it should.

\subsection{3D topological insulators}
We next study the phases of 3D topological insulators.
We consider the following Wilson--Dirac-type tight-binding Hamiltonian \cite{Liu:3DTI,RyuNomura:3DTI};
   \begin{align} \label{eqn:H}
      H = & \sum_{\bm{x}} \sum_{\mu=x,y,z} \left[\frac{{\rm i}t}{2} c^{\dag}_{{\bm{x}}+{\bf e}_\mu} \alpha_{\mu} c_{\bm{x}}
                                         -\frac{m_{2,\mu}}{2}  c^{\dag}_{{\bm{x}}+{\bf e}_\mu} \, \beta c_{\bm{x}} + \rm{H.c.}\right]  \nonumber \\
            & + (m_0+\sum_{\mu=x,y,z} m_{2,\mu})\sum_{\bm{x}} c^{\dag}_{\bm{x}} \, \beta c_{\bm{x}}
            + \sum_{\bm{x}} v_{\bm{x}} c^{\dag}_{\bm{x}} 1_{4} c_{\bm{x}},
   \end{align}
where $c^{\dag}_{\bm{x}}$ and $c_{\bm{x}}$ are four-component creation and annihilation operators on a site $\bm{x}$,
respectively, and
${\bf e}_{\mu}$ is a unit vector in the $\mu$-direction. $\alpha_{\mu}$ and $\beta$ are gamma matrices defined by
   \begin{align} \label{eqn:gammaMat}
      \alpha_{\mu} =\tau_x\otimes\sigma_\mu= \begin{pmatrix}
                         0     & \sigma_\mu \\
                      \sigma_\mu &    0
                   \end{pmatrix}, \ 
      \beta =\tau_z\otimes 1_2= \begin{pmatrix}
                      1_{2} & 0 \\
                      0 & -1_{2}
                   \end{pmatrix}, 
   \end{align}
where $\sigma_{\mu}$ and $\tau_\mu$ are Pauli matrices that act on the spin and orbital degrees of freedom, respectively.
$m_0$ is the mass parameter, and $m_{2,\mu}$ and $t$ are hopping parameters.
The random potential $v_{\bm{x}}$ is uniformly and independently distributed between $[-W/2,W/2]$.
 We set $m_{2,x} =m_{2,y}=1$ as the energy unit.
 The Hamiltonian belongs to the symplectic class for $W>0$.
We impose fixed boundary conditions in the $x$- and $z$-directions.

\begin{figure}[htb]
  \begin{center}
     \begin{tabular}{cc}    
     \begin{minipage}{0.7\hsize}
  \begin{center}
   \includegraphics[width=0.95\textwidth]{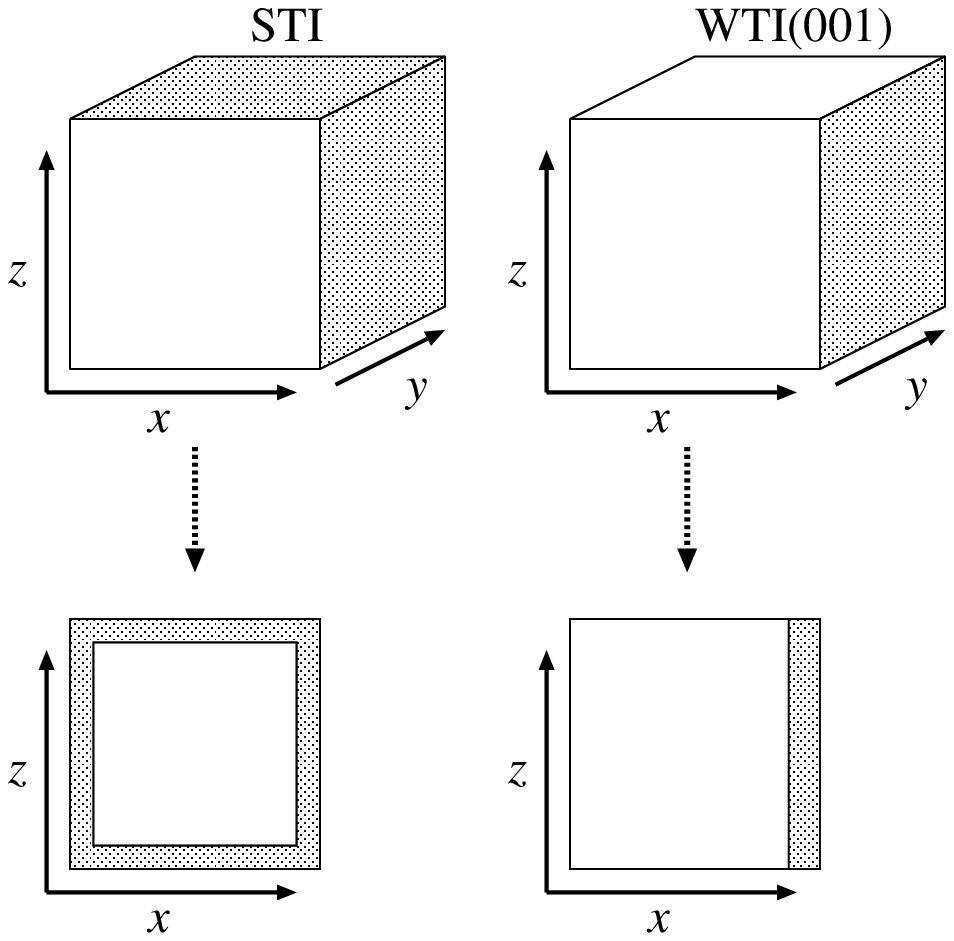}
     \end{center}
 \end{minipage}\\
 
 \hspace{0.5cm}
 \begin{minipage}{0.42\hsize}
  \begin{center}
   \includegraphics[angle=270,width=\textwidth]{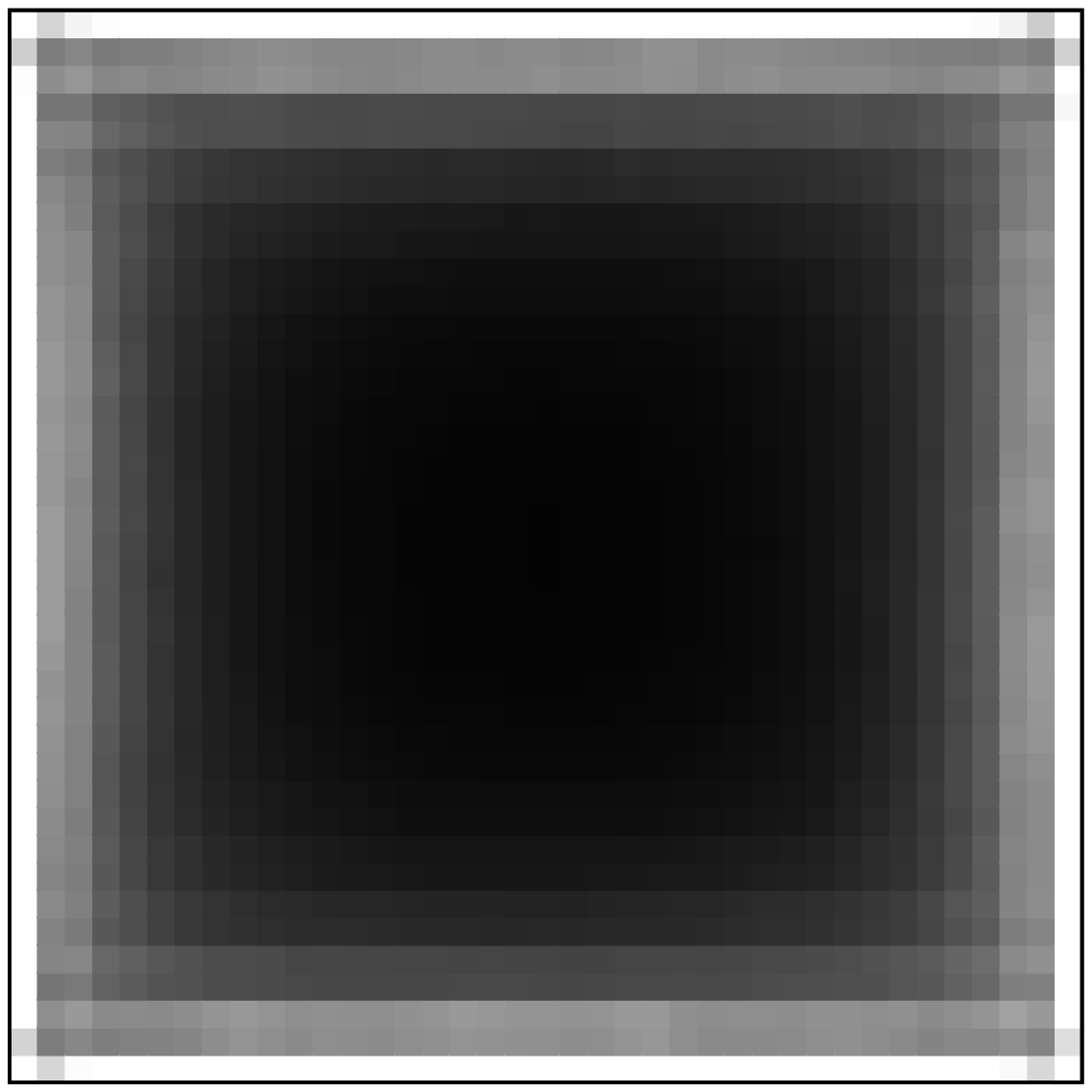}
     \end{center}
 \end{minipage}
 \begin{minipage}{0.42\hsize}
  \begin{center}
  \hspace{-2.5cm}
   \includegraphics[angle=270,width=\textwidth]{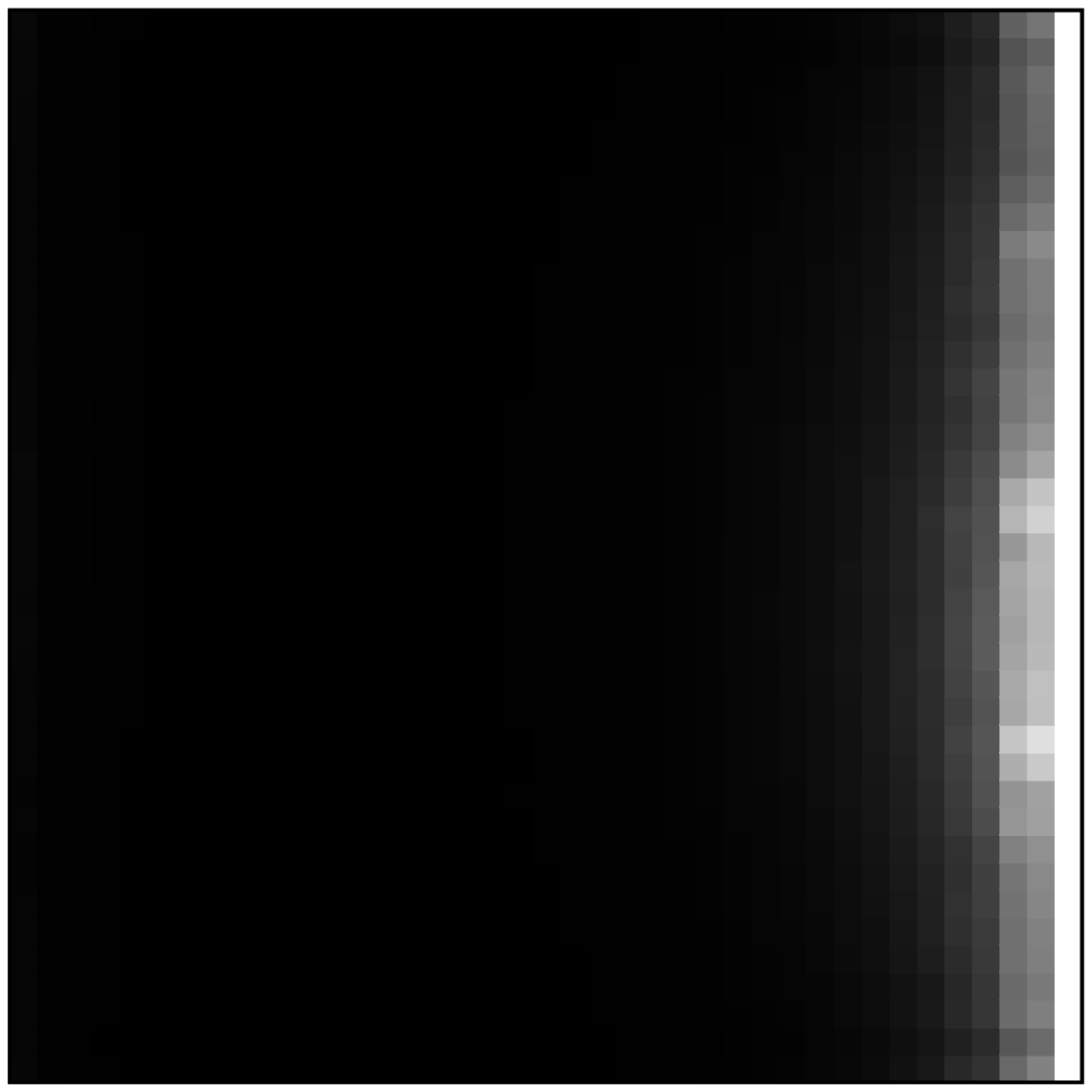}
     \end{center}
 \end{minipage}
  \end{tabular}

 \caption{Schematic of surface states in topological insulators, and the results of integration 
 over the $y$ coordinate of numerically calculated eigenfunctions (bottom panels).  
 The periodic boundary condition is imposed in the
 $y$-direction, and the fixed boundary condition in the $x$- and $z$-directions.
In the case of WTI(001), whether the strong amplitude region appears on the right 
($x\approx 40$) or left ($x\approx 1$)
 depends on the specific configuration of randomness.
In the bottom panels, the modulus squared of a numerically obtained eigenfunction
 integrated over $y$ is shown.
 $(m_0, W)=(-0.16, 3)$ (STI, bottom left) and (-1.16, 3) [WTI(001), bottom right].}
\label{fig:TISchematic}
\end{center}
\end{figure}

We first set $t=2$ and $m_{2,z}=0.5$.  In this case, the
OI phase appears at $m_0>0$,
the STI phase at $0>m_0>-1$,
the WTI phase with weak index (001)  at $-1>m_0>-2$,
and  the WTI phase with weak index (111)  at $-2>m_0>-3$.
\cite{Imura12,Kobayashi15}

Let us first consider the features of surfaces states in the presence of a small ramdomness.
The situation is schematically described in Fig.~\ref{fig:TISchematic}.
In the case of an STI or WTI(111),
at $E\approx 0$, the surface states appear
in the $x$-$y$ and $y$-$z$ planes, but not in the $x$-$z$ plane, since we impose the periodic boundary
condition in the $y$-direction.
After integration over $y$, we expect amplitudes along the sides of the $x$-$z$ plane.
On the other hand,
the surface states at $E\approx 0$ appear only in the  $y$-$z$ planes for WTI(001),
and the large amplitudes are expected along the $z$ side.
Whether they appear on the right  or  left $z$ side depends on the configuration of random potential.

\begin{figure}[htb]
  \begin{center}
     \begin{tabular}{cc}     
 \begin{minipage}{0.48\hsize}
  \begin{center}
   \includegraphics[angle=0,width=0.9\textwidth]{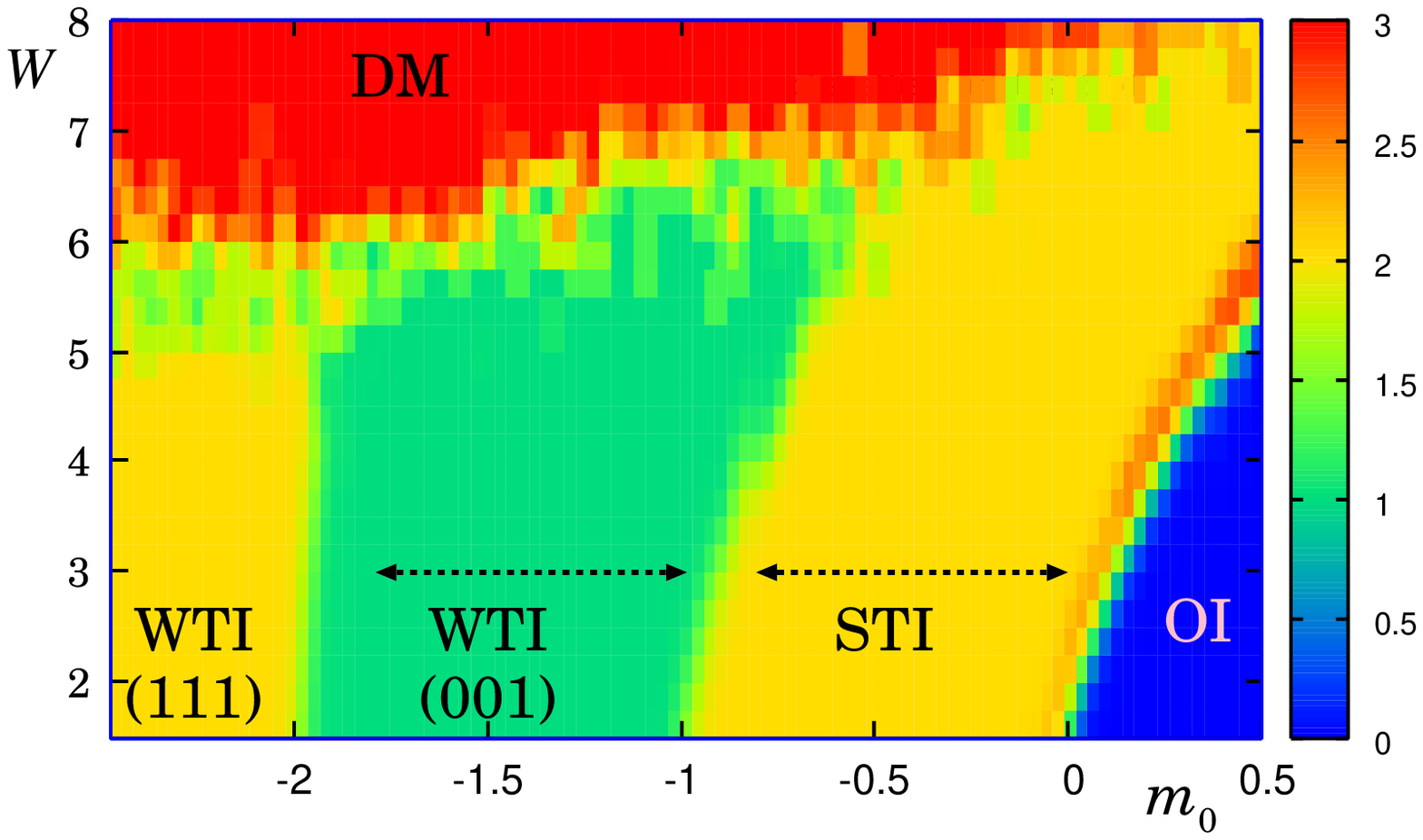}
      \hspace{1.6cm} (a)
     \end{center}
 \end{minipage}
 \begin{minipage}{0.48\hsize}
  \begin{center}
   \includegraphics[angle=0,width=0.9\textwidth]{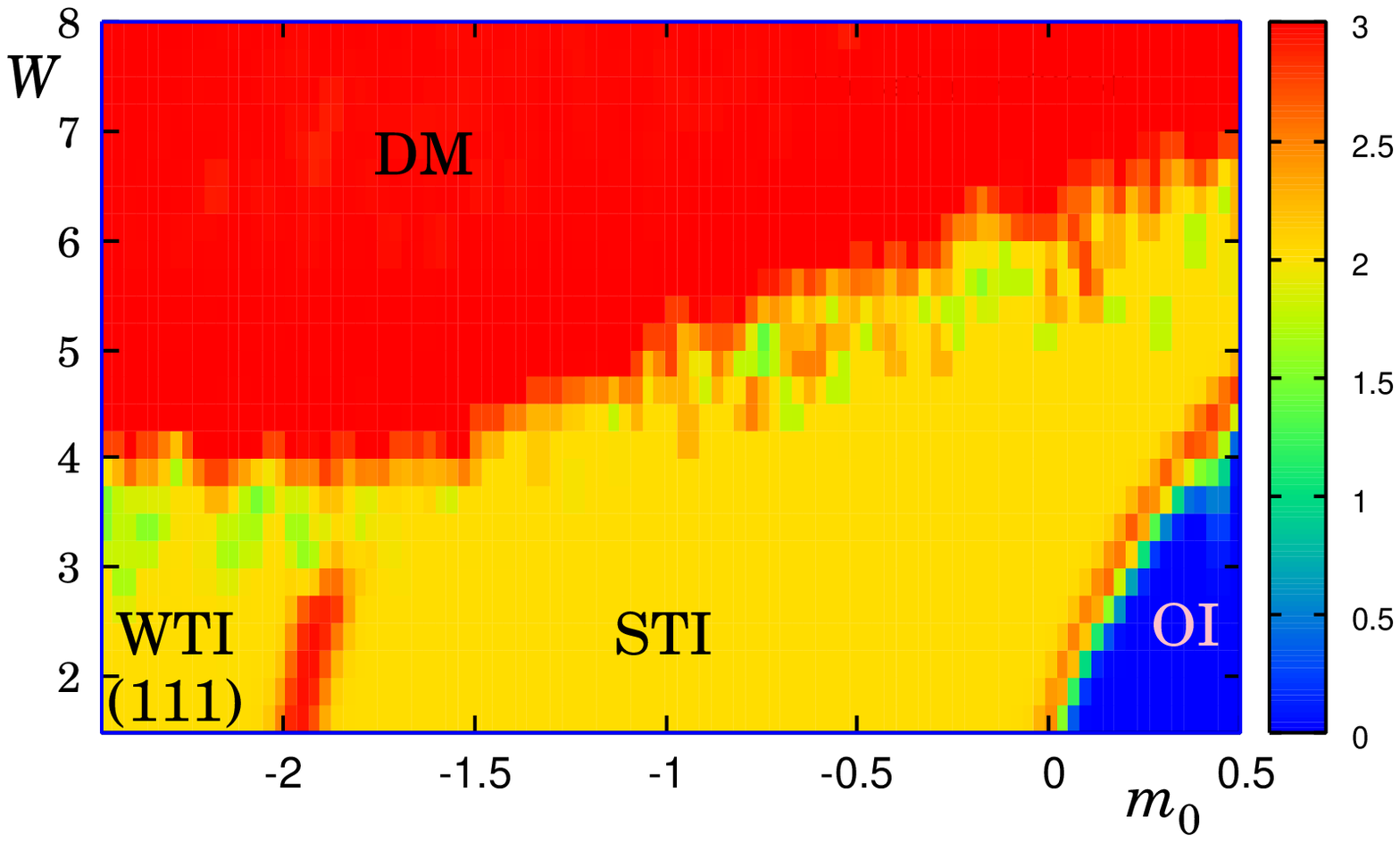}
      \hspace{1.6cm} (b)
     \end{center}
 \end{minipage}
  \end{tabular}

 \caption{(Color) Color map of $P_\mathrm{OI}, P_\mathrm{W001}, P_\mathrm{SW111}$, and $P_\mathrm{DM}$.
 The intensity $0\times P_\mathrm{OI}+ 1\times P_\mathrm{W001} +2\times  P_\mathrm{SW111}+3\times  P_\mathrm{DM}$ is plotted.
 The shifts of the phase boundaries OI/STI, STI/WTI(001),  and 
 WTI(001)/WTI(111) by randomness are clearly seen.  The arrows in (a) indicate the lines along which machine learning for
 STI and WTI(001) has been performed.
 The phase diagram where the transfer matrix method is not applicable, i.e., $t=m_{2,z}=1$,
  is displayed in (b).}
\label{fig:phaseDiagramTI}
\end{center}
\end{figure}

We set $W=3.0$ and
varied $m_0\in [-1.8,-1]$ to teach the features of WTI(001),
and $m_0\in [-0.8,0]$ to teach those of STI.
These training parameters are along lines in the $W$-$m_0$ plane,
which are shown as dotted arrows in Fig.~\ref{fig:phaseDiagramTI}(a).
To teach the features of DM, we set $W=10.0$ and varied $m_0 \in [-2.5,0.5]$,
while, for OI, we set $W=3.0$ and varied $m_0 \in [0.2,0.7]$.
Actually, we do not know the phase diagram for the set of parameters we consider,
so we assumed that  changing $m_{2,z}=1$ to $0.5$ will not change the phase
diagram markedly, and chose the parameters according to the knowledge of the
randomness-free case, $m_{2,z}=0.5$ and $W=0$, together with  information on
the phase diagram for
 the isotropic but disordered case, $m_{2,z}=1$ and $W>0$.\cite{Kobayashi13}

After teaching 4000 eigenfunctions in each phase, we  prepared
100$\times$27 eigenfunctions with different $m_0$ (100 values) and $W$ (27 values),
and let the machine determine which phase each eigenfunction belongs to.
We calculate
the probabilities $P_\mathrm{OI}, P_\mathrm{W001}, P_\mathrm{SW111}, $ and
 $P_\mathrm{DM}(=1-P_\mathrm{OI}-P_\mathrm{W001}-P_\mathrm{SW111})$
that a given eigenfunction belongs to OI, WTI(001), STI or WTI(111), and DM, respectively.
Note that the present method cannot distinguish WTI(111) from STI,
since WTI(111) contains surface states on all surfaces normal to the $x$- and $z$-axes just as in STI,
but, from the knowledge of the randomness-free case, we can reasonably determine whether the phase
is STI or WTI(111).

The probabilities of OI, WTI(001), STI or WTI(111), and DM
are displayed as
a color map in the $W$-$m_0$ plane [Fig.~\ref{fig:phaseDiagramTI}(a)].
We see that the phase boundaries between insulators with different topologies
shift as we increase $W$.
For example, when we start with the OI phase, say $(m_0,W)= (0.3, 1.5)$ and increase the disorder $W$,
we enter into the STI phase at $W\approx 6$.  This is called the topological Anderson insulator (TAI) transition.\cite{Li09a,Groth09,Guo10}
The present method captures TAI and gives a phase diagram quantitatively consistent with
that obtained via the transfer matrix method.\cite{Kobayashi13,KobayashiPrivateComm}
It should be emphasized that one-dimensional training along a few finite lines in a parameter space enables us to draw the
2D phase diagram.

We next apply the trained CNN of the above case, $t=2$ and  $m_{2,z}=0.5$, to
the case of  $t=m_{2,z}=m_{2,x}=m_{2,y}(=m_{2})$.
In the absence of randomness, this choice of
parameters gives OI for $m_0>0$, STI for $0>m_0>-2$, and WTI(111) for $-2>m_0>-4$.\cite{Imura12}
The standard method of using the transfer matrix\cite{Kobayashi13} to determine the phase diagram
in the presence of disorder
breaks down for this choice of parameters, since the transfer matrix connecting
a layer to the next layer is not invertible for $t^2-m_2^2=0$.\cite{RyuNomura:3DTI}
This choice of parameters, therefore, demonstrates the wider validity of the machine learning method.
Results are shown in Fig.~\ref{fig:phaseDiagramTI}(b).

Note that,
 although the colors of STI and WTI(111) are the same, we can distinguish them by the emergence of the metal phase
(in fact, it is a Dirac semimetal phase\cite{Kobayashi14})
sandwiched between insulators with different topologies.
Whether they are STI or WTI(111) can be determined
from the randomness-free limit.
Note also that the green region that appears on the OI/STI boundary is not the WTI(001) phase, but
an artifact of the color map.
See Appendix for details.

\subsection{Layered Chern insulator}
We next consider two-dimensional
Chern insulators\cite{Dahlhaus11,Liu16,Chang16}
and stack them in the $z$-direction to form a three-dimensional
Chern insulator and a Weyl semimetal.\cite{Liu16,Yoshimura16}
We begin with a spinless
two-orbital tight-binding model on a square lattice,
which consists of an $s$-orbital and a $p\equiv p_x+ip_y$ orbital,~\cite{Qi08}
and stack them in the $z$-direction to form a cubic lattice,
\begin{align}
H = & \sum_{{\bm x}} \left(
(\epsilon_s + v_s({\bm x})) c^{\dagger}_{{\bm x},s} c_{{\bm x},s}
+ (\epsilon_p + v_p({\bm x})) c^{\dagger}_{{\bm x},p} c_{{\bm x},p}\right)   \nonumber \\
 +& \sum_{{\bm x}}\Big(-\sum_{\mu=x,y} (
t_s c^{\dagger}_{{\bm x} + {\bm e}_{\mu},s} c_{{\bm x},s}
- t_p c^{\dagger}_{{\bm x} + {\bm e}_{\mu},p} c_{{\bm x},p}) \nonumber \\
& +  t_{sp}
(c^{\dagger}_{{\bm x}+{\bm e}_x,p}
- c^{\dagger}_{{\bm x} - {\bm e}_x,p})  \!\ c_{{\bm x},s} 
-  it_{sp}
(c^{\dagger}_{{\bm x}+{\bm e}_y,p}
- c^{\dagger}_{{\bm x} - {\bm e}_y,p})  \!\ c_{{\bm x},s}
+{\rm H.c.}\Big)  \nonumber \\
-& \sum_{{\bm x}} \Big( t^{\prime}_s c^{\dagger}_{{\bm x} + {\bm e}_{z},s} c_{{\bm x},s}
+ t^{\prime}_p c^{\dagger}_{{\bm x} + {\bm e}_{z},p} c_{{\bm x},p} + {\rm H.c.} \Big)
\,,\nonumber 
\label{tb1}
\end{align}
where $\epsilon_s$,  $v_s({\bm x})$, $\epsilon_p$, and $v_p({\bm x})$ 
denote the atomic energies and disorder potentials for the $s$- and $p$-orbitals, respectively.
Both $v_s({\bm x})$ and $v_p({\bm x})$ 
are uniformly distributed within $[-W/2,W/2]$ with
an independent probability distribution. $t_s$, $t_p$, and $t_{sp}$ are
 transfer integrals between neighboring $s$-orbitals, $p$-orbitals, and that between
$s$- and $p$-orbitals, respectively.
$t^{\prime}_{s}$ and $ t^{\prime}_{p}$ are 
interlayer transfer integrals.

As in Ref.\cite{Liu16} we set $\epsilon_s-\epsilon_p=-2(t_s+t_p)$,
$t^{\prime}_s=-t^{\prime}_p>0$, $t_s=t_p>0$, and $t_{sp}=4t_s/3$,
and took $4t_s$ as the energy unit.
In the absence of randomness,
this set of parameters realizes a CI with a large band gap in the 2D limit,
$\beta \equiv \frac{t^{\prime}_p-t^{\prime}_s}{2(t_s+t_p)} =0$.
As long as $1/2> |\beta|\ge 0$, the system is fully gapped, and the system
belongs to the CI phase. The system enters into the 3D WSM phase for $|\beta|>1/2$.\cite{Liu16}

In the presence of randomness, there are four phases: CI, WSM, DM, and AI.
AI phase appears in the large $W$ region.
Here, we focus on the first three phases by considering $W<4.5$ and $0.3<\beta<0.6$.
Note that there is a correspondence between CI and WTI, and WSM and STI.\cite{Yoshimura16}

\begin{figure}[htbp]
  \begin{center}
     \begin{tabular}{cc}    
     \begin{minipage}{0.7\hsize}
  \begin{center}
   \includegraphics[width=0.95\textwidth]{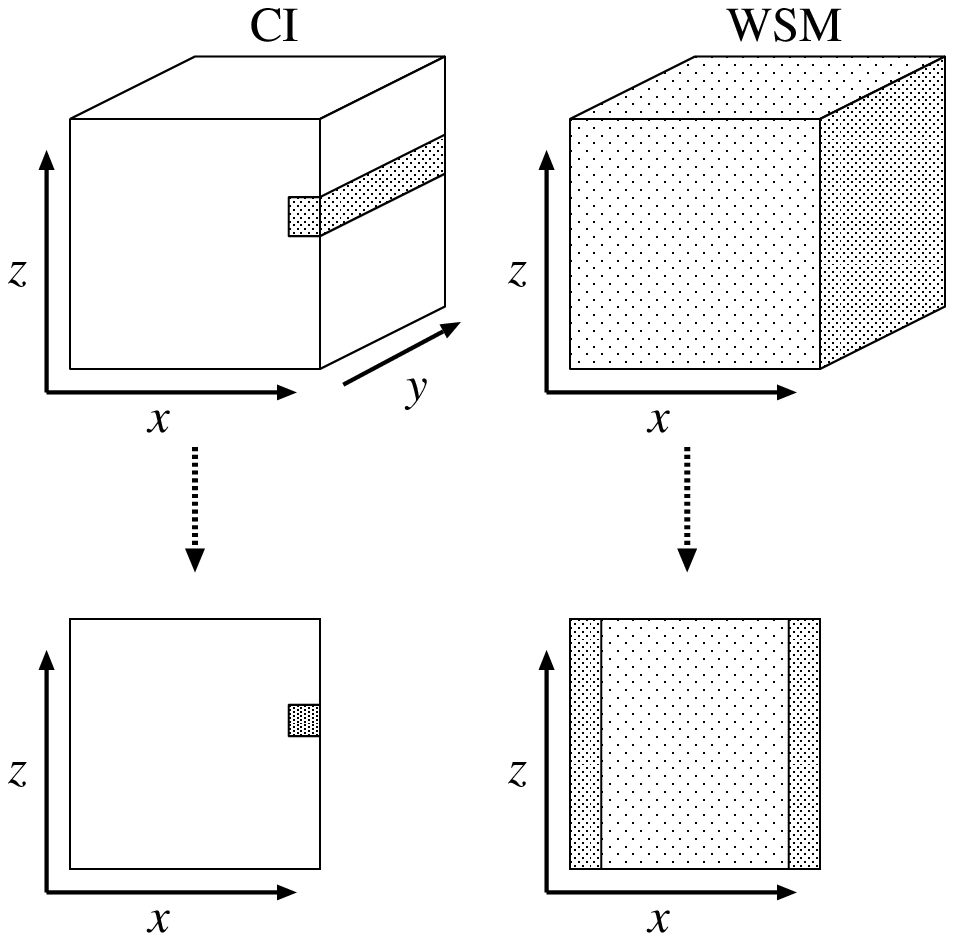}
     \end{center}
 \end{minipage}\\

   \hspace{-0.6cm}
 \begin{minipage}{0.38\hsize}
  \begin{center}
   \includegraphics[angle=90,width=\textwidth]{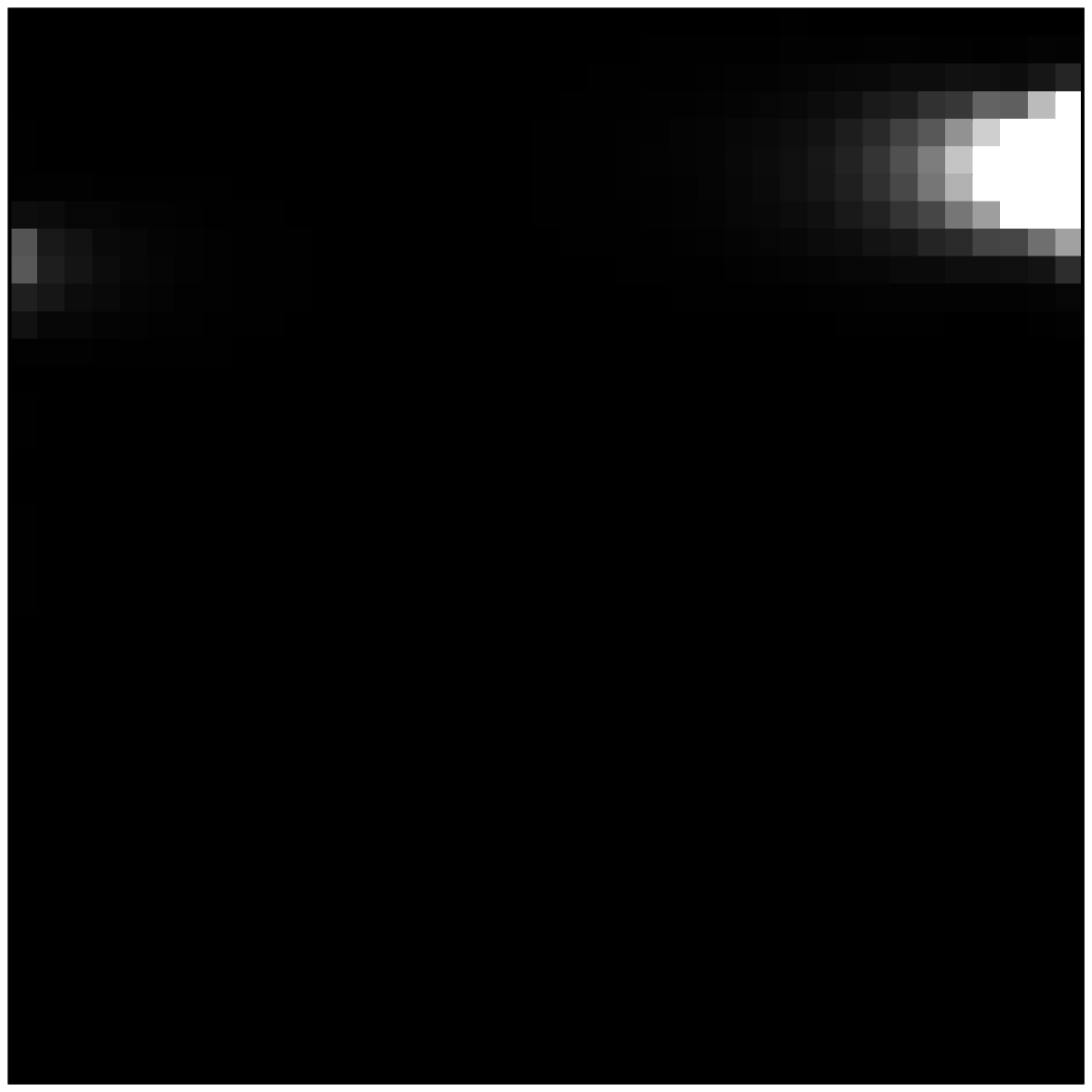}
     \end{center}
 \end{minipage}

  \hspace{-0.7cm}
 \begin{minipage}{0.38\hsize}
  \begin{center}
  \vspace{-0.2cm}
   \includegraphics[angle=270,width=\textwidth]{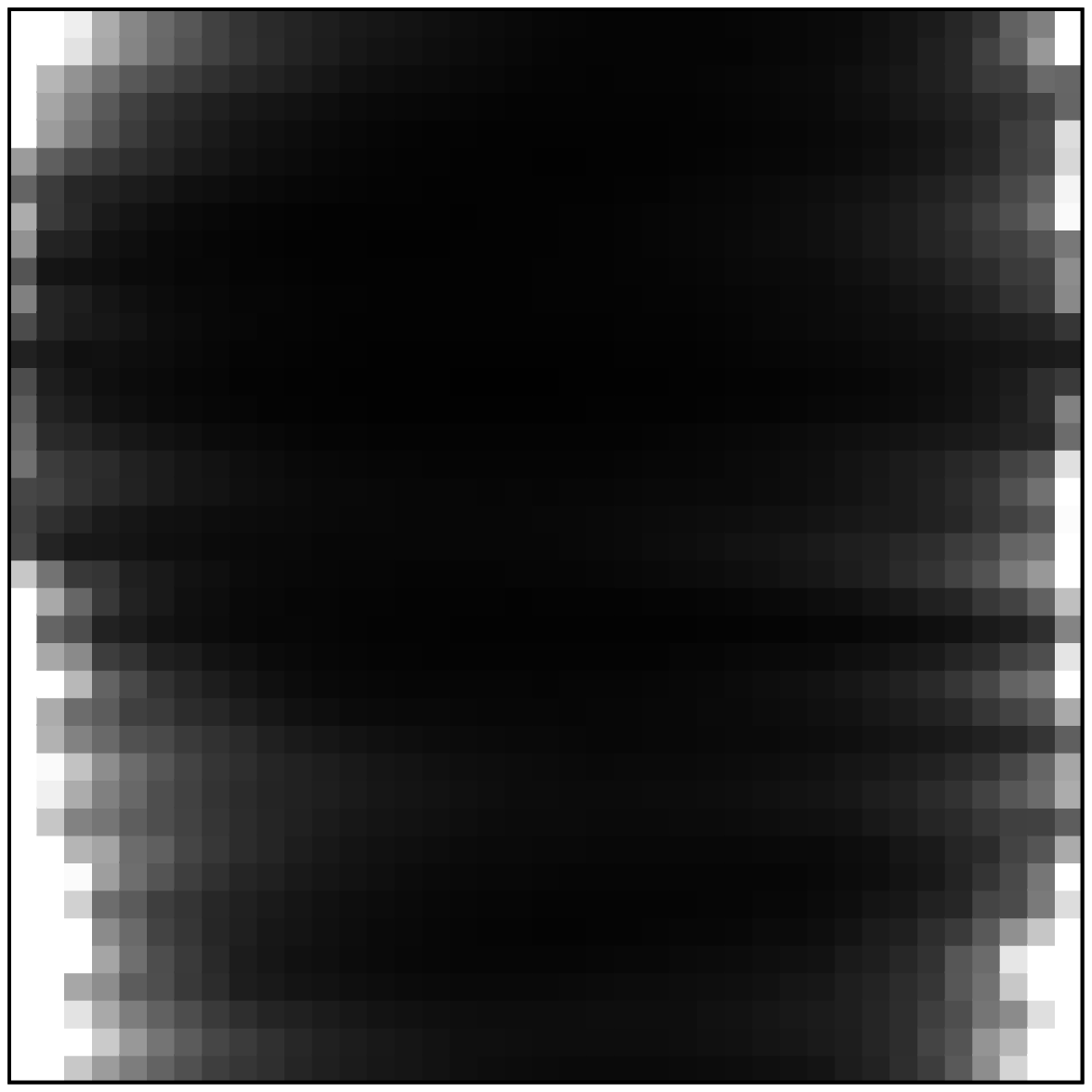}
     \end{center}
 \end{minipage}
  \end{tabular}

 \caption{Schematic of amplitudes and the results of integration 
 over the $y$ coordinate of numerically calculated eigenfunctions (bottom panels).
 The periodic boundary condition is imposed in the
 $y$- and $z$-directions, and the fixed boundary condition in the $x$-direction.
 In the bottom panels, the modulus squared of a numerically obtained eigenfunction
 integrated over $y$ is shown.
$(\beta, W)$=$(0.4,0.6)$ (CI,  bottom left) and  $(\beta, W)$=$(0.55,0.9)$ (WSM,  bottom right).}
\label{fig:WSMSchematic}
\end{center}
\end{figure}

As in the previous subsection, let us first consider the features of states near $E=0$ in the case of  a small randomness.
The situation is schematically described in Fig.~\ref{fig:WSMSchematic},
where periodic boundary conditions are imposed in the $y$- and $z$-directions,
while the fixed boundary condition is imposed in the $x$-direction.
In the case of a CI,
at $E\approx 0$, edge states run along the $y$-direction, and
after integration over $y$, we expect amplitudes in a dot in the $x$-$z$ plane
close to a side in the $z$-direction.
On the other hand,  in WSM,
the surface states corresponding to the Fermi arc\cite{Okugawa14} appear on surfaces normal to the $x$-direction.
Hence, large amplitudes are expected along $z$ sides after integration along $y$.
Owing to the presence of the bulk Weyl node (shown as dilute dots) near the same energy $E=0$,
the right and left surface states are coupled
and the high-amplitude regions appear on both the right and left sides.

\begin{figure}[htbp]
  \begin{center}
   \includegraphics[angle=0,width=0.40\textwidth]{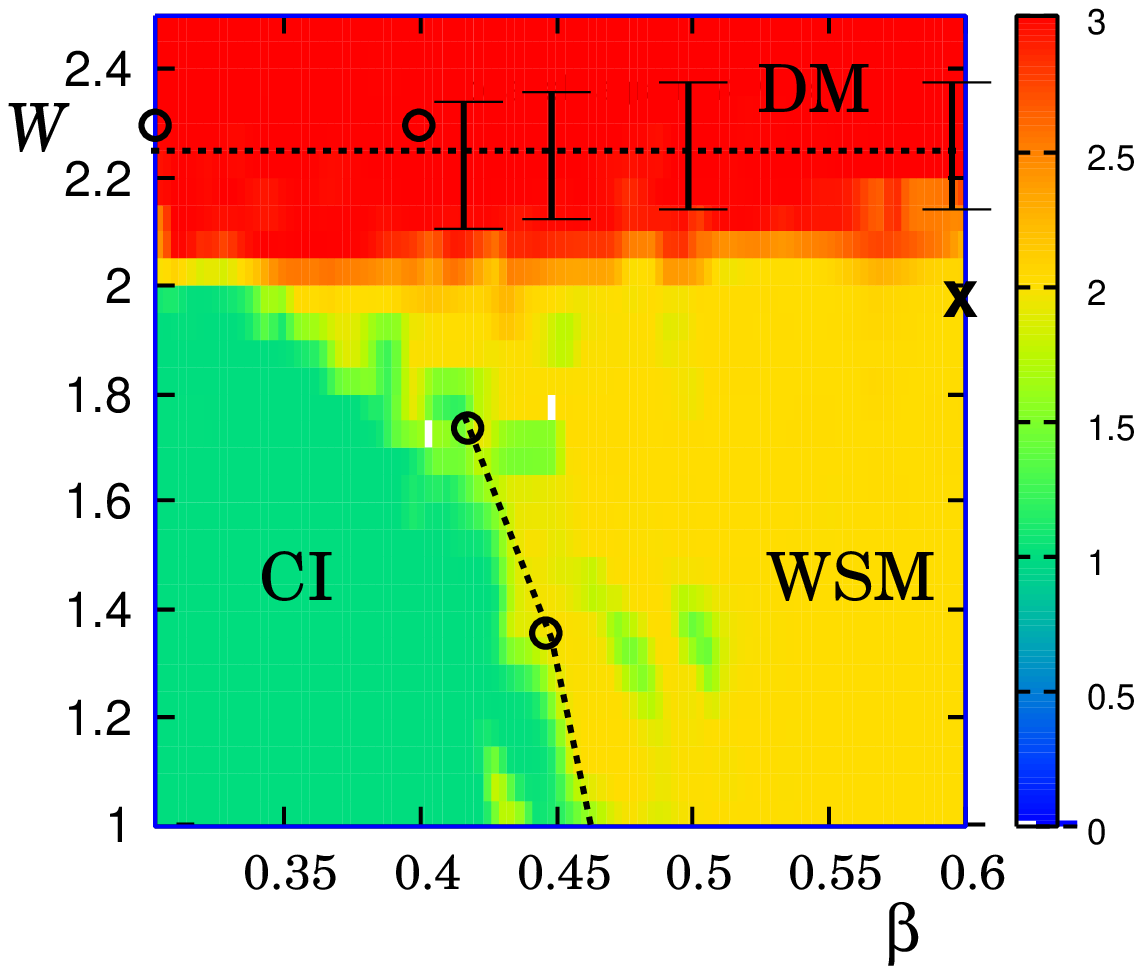}
 \caption{(Color) Color map of $P_\mathrm{CI}, P_\mathrm{WSM},$ and  $P_\mathrm{DM}$.
 The intensity $1\times P_\mathrm{CI}+2\times P_\mathrm{WSM}+ 3\times P_\mathrm{DM}$ is plotted.
Bars with errors and circles ($\circ$) indicate the transfer matrix estimate of the
critical points,\cite{Liu16} and dotted lines are a guide to the eye.
 The cross ($\times$) at $(\beta, W)\approx (0.6,2.0)$ indicates the WSM/DM phase boundary estimated by
 the scaling of density of states.\cite{Kobayashi14,Liu16}}
\label{fig:phaseDiagramWSM}
\end{center}
\end{figure}

We set $\beta=0.4$ and
varied $W\in [2.5,4.5]$ to teach the features of DM, and
$W\in [0.4,1.6]$ to teach those of CI.  For WSM, we set $\beta=0.55$ and varied $W \in [0.3,1.5]$.
For DM and WSM, we have prepared 3000 samples for teaching the features of eigenfunctions,
while 4000 samples for teaching the features of CI.
We then varied $\beta$ and $W$ and let the machine calculate
the probabilities $P_\mathrm{CI}, P_\mathrm{WSM}$, and $P_\mathrm{DM}(=1-P_\mathrm{CI}-P_\mathrm{WSM})$ that a given eigenfunction belongs to CI, WSM, and DM,
respectively.
We then draw 
a color map in the $W$-$\beta$ plane as shown in Fig.~\ref{fig:phaseDiagramWSM},
which qualitatively reproduces the phase diagram obtained by the transfer matrix method.\cite{Liu16}

\section{Summary and Concluding Remarks}
In this paper, we have considered three examples of 3D
quantum phase transitions: the Anderson transition,
transitions in topological insulators, and transitions in Weyl semimetal. 
From the universality class point of view, the Anderson transition belongs to
the orthogonal class, while topological insulators belong to the symplectic class.
The Weyl semimetal realized by stacking Chern insulators belongs to the unitary class.
We therefore have studied Wigner--Dyson classes.\cite{Wigner51,Dyson61,Dyson62}
Applications of this method to other 7 nonstandard universality classes with chiral and/or particle--hole
symmetries \cite{Zirnbauer96,altland97} would be interesting.

By machine learning, we could derive  approximate phase diagrams.
In particular, the shift of phase boundaries between insulators with different topologies
as we increase the disorder strength $W$ has been captured.
It should be emphasized that the phase diagrams
of topological insulators and Weyl semimetals have been determined by the results of
learning along finite one-dimensional lines in a two-dimensional parameter space
[see, for example, the arrows in Fig.~\ref{fig:phaseDiagramTI}(a)].


The phase boundaries between the DM and the
 insulating phases such as OI, STI, and WTI, on the other hand, are not as sharply determined as those of 
 the insulating phases with different topologies.
 One possible reason is that the fluctuation of the wave function amplitude in the 
 DM phase is much larger than that of the topological surface states.
 A significant fluctuation of the wave function also arises in the case of the WSM phase,
 where bulk wave functions near the Weyl nodes and surface states forming the Fermi arc are coupled.  This might be the reason why the phase boundary between CI and WSM is
obscurer than those of OI/STI, STI/WTI, and WTI(001)/WTI(111).

In some systems, the transfer matrix approach is not applicable.  In this paper,
we have demonstrated [Fig.~\ref{fig:phaseDiagramTI}(b)] that the deep
learning method proposed here can
be applied to systems where the transfer matrix is not defined.  Other applications
to complicated lattices such as quantum percolation\cite{Avishai92,Berkovits96,Kaneko99,Ujfalusi14} 
and fractal lattice\cite{Asada06} are important future topics.

For the analysis of the Anderson transition, we have used the known value of the critical disorder $W_c$
for training CNN. A natural question arises whether we can determine the critical disorder $W_c\approx 16.54$
by the current method instead of by other methods.
One possibility for detecting the critical disorder might be to prepare the Fourier-transformed wave functions.
After Fourier transformation, localized states become delocalized in the $k$-space while delocalized ones become localized.
This duality is broken if we estimate $W_c$ wrongly.
Careful monitoring of the validation error of CNN for real-space and $k$-space wave functions
would enable us to detect the wrong choice of $W_c$.


Machine learning has recently been applied to solving several problems of condensed matter physics such as
Ising and spin ice models\cite{Carrasquilla16,Tanaka16}, 2D topological system\cite{Zhang16},
and strongly correlated systems.\cite{Carleo16,Broecker16,Chng16,Li16,Nieuwenburg16,Huang17}
Applications to interacting electron systems with disorder\cite{Mills17} such as the Anderson--Hubbard model\cite{Shinaoka09,Harashima14}
as well as to topological systems of dirty bosons such as quantum magnon Hall insulators\cite{Xu16}
and topological photonic insulators\cite{Hu15}
are interesting problems to be studied in the future.

\begin{acknowledgments}
The authors would like to thank Koji Kobayashi for useful comments and for showing us a phase diagram
corresponding to Fig.~\ref{fig:phaseDiagramTI}(a) obtained via the transfer matrix method
prior to publication.
Tomi Ohtsuki thanks  Keith Slevin, Ken-Ichiro Imura, Shang Liu, and Ryuichi Shindou for fruitful
collaborations on 3D random electron systems treated in this paper.
This work was partly supported by JSPS KAKENHI Grant No. JP15H03700.
\end{acknowledgments}

\newpage
\noindent
{\bf Appendix}

\begin{figure}[htbp]
  \begin{center}
    \begin{tabular}{c}
    
     \begin{minipage}{0.46\hsize}
  \begin{center}
   \includegraphics[width=70mm,angle=0]{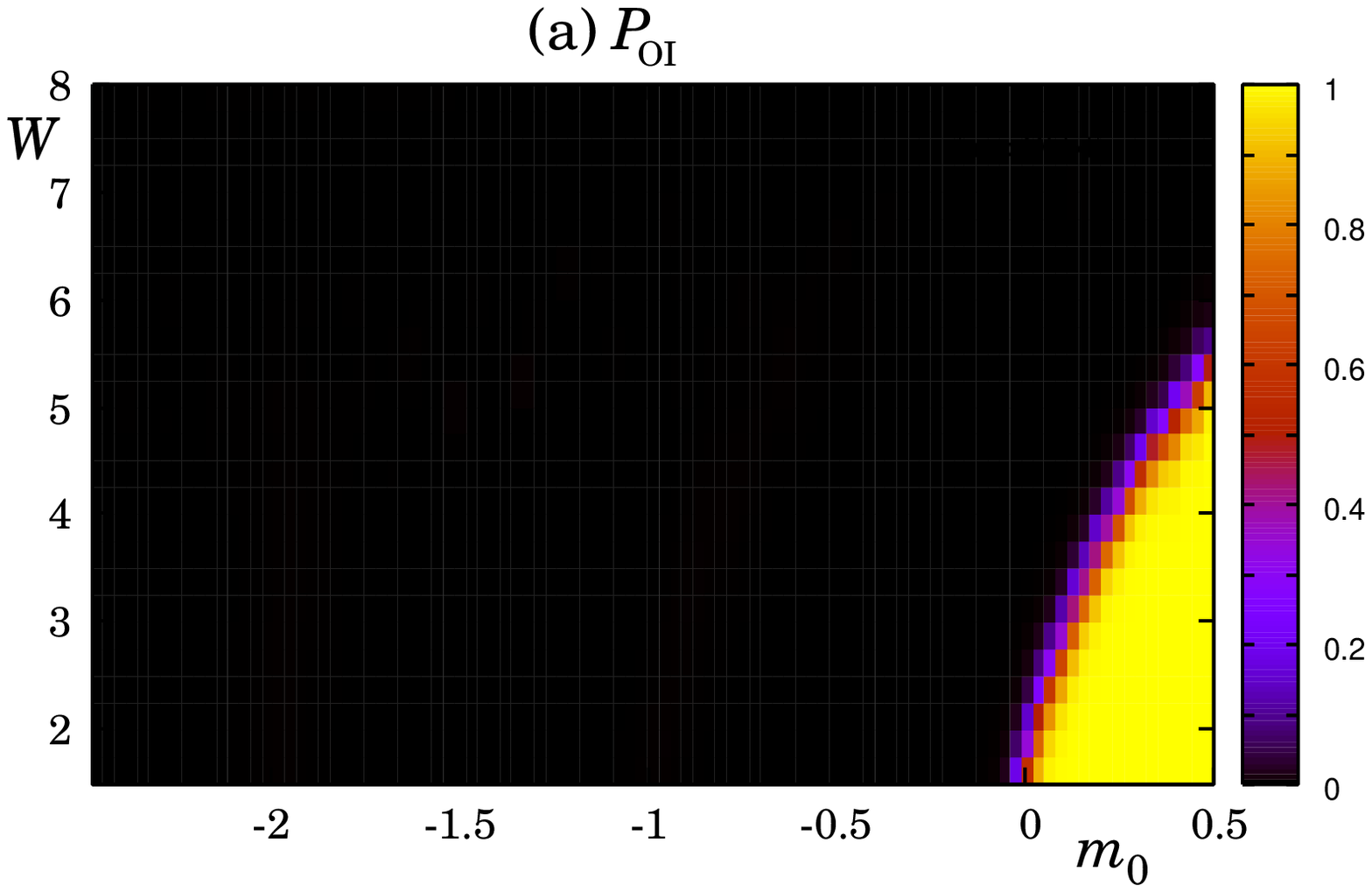}
     \end{center}
 \end{minipage}
 
 \begin{minipage}{0.46\hsize}
  \begin{center}
   \includegraphics[width=70mm,angle=0]{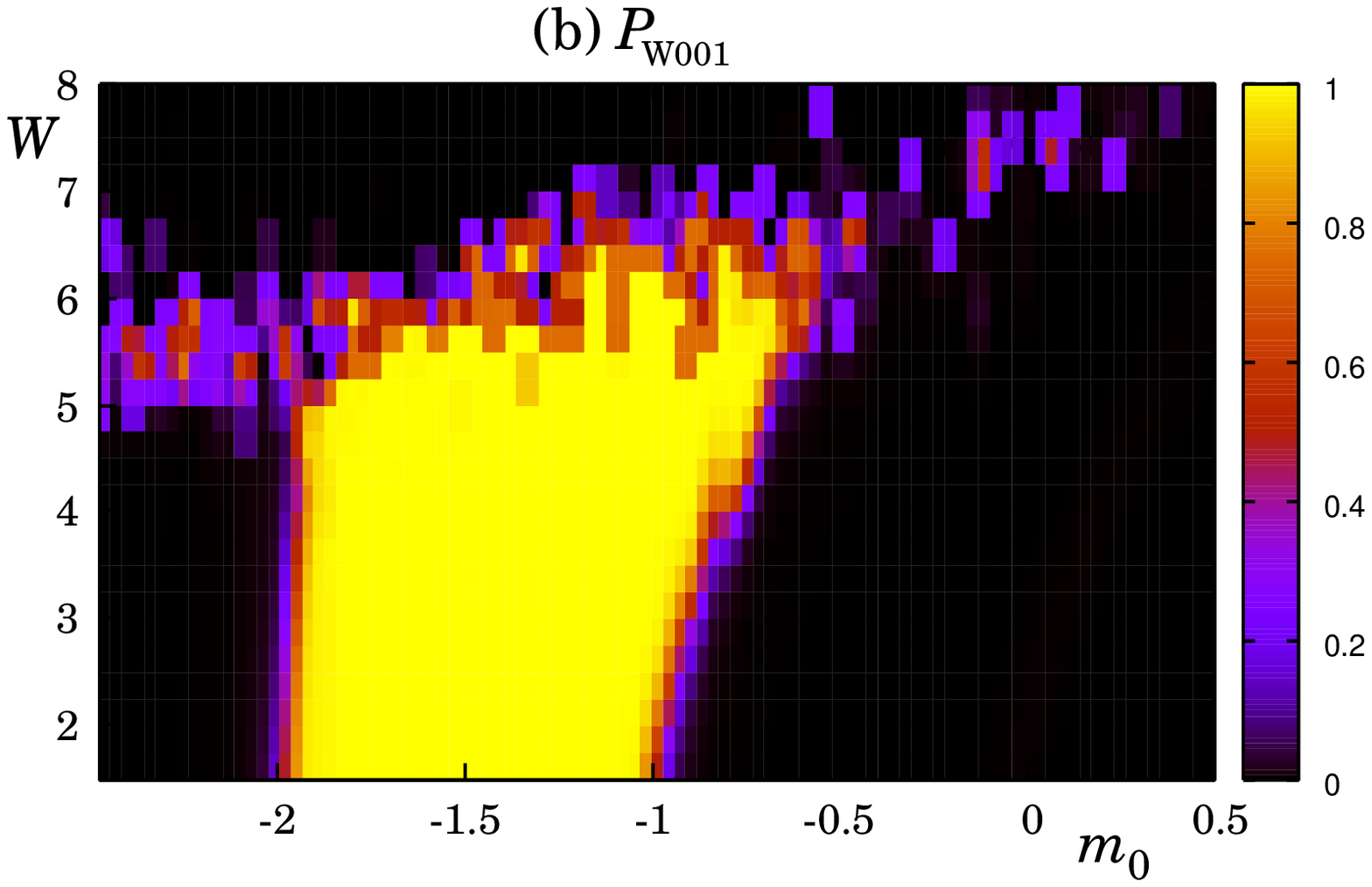}
     \end{center}
 \end{minipage}
 
 \\

 \begin{minipage}{0.46\hsize}
  \begin{center}
   \includegraphics[width=70mm,angle=0]{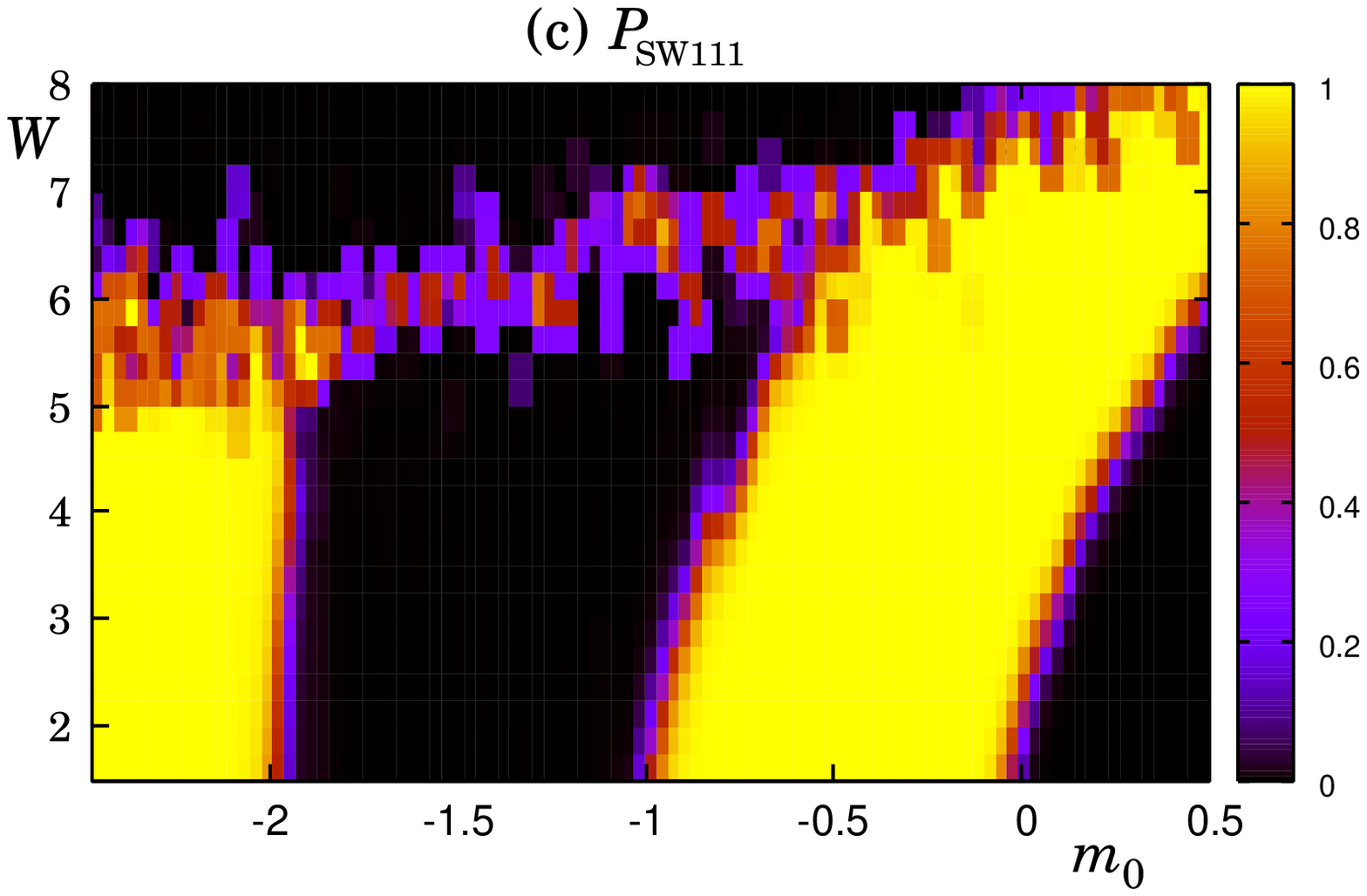}
  \end{center}
 \end{minipage}

 \begin{minipage}{0.46\hsize}
  \begin{center}
   \includegraphics[width=70mm,angle=0]{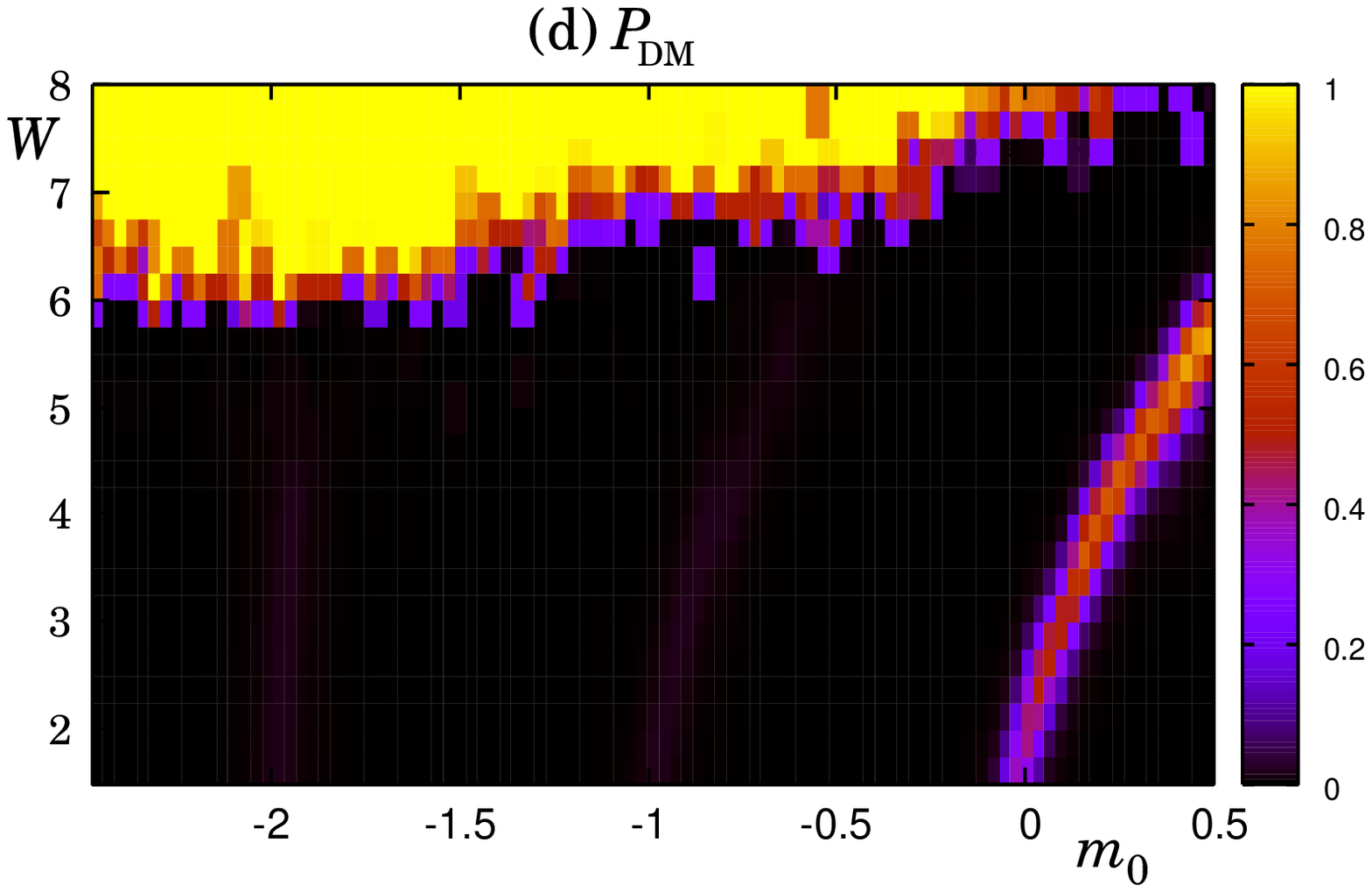}
  \end{center}
 \end{minipage}
 
  \end{tabular}
 \caption{(Color) 
 Map view of $P_i$ ($i=$ OI, W001, SW111, DM) corresponding to  Fig.~\ref{fig:phaseDiagramTI} (a).
 (a) OI, (b) WTI(001), (c) STI/WTI(111), and (d) DM.
 }
\label{fig:mapEachPhase}
\end{center}
\end{figure}

In the main text, we have plotted $0\times P_\mathrm{OI}+ 1\times P_\mathrm{W001} +2\times  P_\mathrm{SW111}+3\times  P_\mathrm{DM}$,
which is a convenient way to see various phases in a single figure.
Around the phase boundaries where $P_i$ ($i=$ OI, W001, SW111, DM) deviates from 0 or 1, however, this quantity becomes
ambiguous.  In this Appendix, we show each $P_i$ to complement the plot in the main text, taking the case of
Fig.~\ref{fig:phaseDiagramTI}(a) as an example.
From Fig.~\ref{fig:mapEachPhase}(b), we see that no WTI(001) phase appears near the OI/STI boundary, so the
green region near OI/STI boundary in Fig.~\ref{fig:phaseDiagramTI}(a) is not the
 WTI(001) phase, but is a transient region towards the
metal (Dirac semimetal, to be precise) phase, which appears at the phase boundary
 [see Figs.~\ref{fig:mapEachPhase}(a) and \ref{fig:mapEachPhase}(d)].

\newpage

\bibliography{tomokiJPSJ2017}

\end{document}